\documentstyle[emulateapj, psfig]{article}
\newcommand \Lya          {\hbox{Ly$\alpha$}}
\newcommand \Zsun          {\hbox{Z$_{\odot}$}}
\newcommand \Msun          {\hbox{M$_{\odot}$}}
\begin{document}
\title{The Metallicity of the Redshift 4.16 Quasar BR2248-1242$^\ast$}
\author{Craig Warner\altaffilmark{1}, Fred Hamann\altaffilmark{1}, 
Joseph C. Shields\altaffilmark{2}, Anca Constantin\altaffilmark{2},
Craig B. Foltz\altaffilmark{3}, \& Frederic H. Chaffee\altaffilmark{4}}
\altaffiltext{1}{Department of Astronomy, University of Florida,
211 Bryant Space Science Center, Gainesville, FL 32611-2055,
\\E-mail: warner@astro.ufl.edu, hamann@astro.ufl.edu
Webpage: http://www.astro.ufl.edu/$\sim$warner/BR2248/}
\altaffiltext{2}{Department of Physics and Astronomy, Ohio University,
Athens, OH 45701}
\altaffiltext{3}{MMT Observatory, Tucson, AZ 85721}
\altaffiltext{4}{W. M. Keck Observatory, Kamuela, HI 96743
\\$^\ast$Observations reported here were obtained at the MMT Observatory, a
joint facility of the Smithsonian Institution and the University of Arizona.}
\begin{abstract}

	We estimate the metallicity in the broad emission-line region of
the redshift $z=4.16$ quasar, BR2248-1242, by comparing line ratios
involving nitrogen to theoretical predictions.  BR2248-1242 
has unusually narrow emission lines with large equivalent widths, thus
providing a rare opportunity to measure several line-ratio abundance
diagnostics.  The combined diagnostics indicate a metallicity of $\sim$2
times solar.  This result suggests that an episode of vigorous
star formation occurred near BR2248-1242 prior to the observed $z=4.16$
epoch.  The time available for this enrichment episode is only $\sim$1.5 Gyr
at $z=4.16$ (for H$_{0}=65$ km s$^{-1}$ Mpc$^{-1}$, $\Omega_{m}=0.3$ and 
$\Omega_{\Lambda} \lesssim 1$).  This evidence for high metallicities
and rapid star formation is consistent with the expected early-epoch evolution
of dense galactic nuclei. 

\end{abstract}

\section{Introduction}

	The prominent emission line spectra of quasars provide valuable
information on the physical state and chemical composition of the gas close
to the quasars.  If this gas was processed by stars in the surrounding host
galaxy then the abundances we measure from the emission lines are a diagnostic
of the star formation history in these environments (see Hamann \& Ferland
1999 for a general review).  At redshifts $z\gtrsim4$, when the universe was
only $\sim$1 to 2 Gyr old (depending on cosmological models, see Figure 1 in
Hamann \& Ferland 1999), the results are especially interesting because they
provide information about very early star formation, galaxy evolution,
and chemical enrichment (Ostriker \& Gnedin 1997; Friaca \& Terlevich 1998).

	Ratios of emission lines involving nitrogen, N, are valuable in
determining the metallicity, $Z$, because of the expected ``secondary" N
production via the CNO cycle (Shields 1976).  The nitrogen abundance
therefore scales as $Z^{2}$ = (O/H)$^{2}$ and N/O scales as $Z$ = O/H,
providing a sensitive metallicity diagnostic even when direct measures of
$Z$ = O/H are not available (Shields 1976; Ferland et al. 1996; Hamann \&
Ferland 1992, 1993, 1999).  Observations of \ion{H}{2} regions
indicate that secondary nitrogen production dominates for $Z\gtrsim0.2$ \Zsun
(Van Zee et al. 1998; Vila-Costas \& Edmunds 1993).  When secondary nitrogen
production dominates, the N abundance is given by: 
\begin{equation}
[{\rm N}/{\rm O}]\ \approx\ [{\rm O}/{\rm H}] - q\ \approx\ \log(Z/\Zsun) - q
\end{equation}
where the square brackets indicate logarithmic ratios relative to solar
(Hamann et al. 2001).  In environments with high star formation rates and
short enrichment times, chemical evolution models predict that
nitrogen production should be delayed relative to oxygen and carbon.  In
Equation 1, $q$ is the logarithmic offset caused by this delay.  It is
believed to range from $\sim$0.0--0.1 for ``slow" chemical evolution, such as
in \ion{H}{2} regions, to $\sim$0.2--0.5 for rapid evolution, such as may
occur in massive galactic nuclei (Hamann et al. 2001).  

	Studies of both emission and absorption lines in quasars suggest
that their metallicities are near or above the solar value (Dietrich et al.
1999, 2000; Osmer et al. 1994; Constantin et al. 2001; Petitjean et al. 1994;
Hamann 1997; Hamann \& Ferland 1999; Hamann et al. 2001).  These metallicities 
are consistent with both observational studies (Pettini 2001) and theoretical
simulations (Cen \& Ostriker 1999) showing that metal abundances at
any time in the universe are dependent on the local density.  At every epoch,
higher density regions, such as the central regions of galaxies, where quasars
reside, should have much higher metallicities than lower density regions such
as galactic halos or the intergalactic medium.  This dependence of metallicity
on density seems to be stronger than any relationship to age (i.e. old does
not necessarily mean metal poor).  Evidence that density affects metallicity
much more than age is found even in our own galaxy, where the central bulge
contains many old but metal rich stars (McWilliam \& Rich 1994;
Idiart et al. 1996).

	Much of the previous emission line analysis relied on the line ratios
of \ion{N}{5} $\lambda 1240$/\ion{He}{2} $\lambda 1640$ and
\ion{N}{5}/\ion{C}{4} $\lambda 1549$.  Ratios of various weak intercombination
lines can be used to test that analysis.  In this paper, we measure the
emission lines in the rest frame UV spectrum of the high redshift ($z=4.16$)
quasar BR2248-1242 (Storrie-Lombardi et al. 1996), and we examine line ratios
involving nitrogen to estimate the metallicity.  We selected BR2248-1242 from
a sample of 44 $z \gtrsim 4$ quasars in Constantin et al. (2001) because it
has unusually narrow emission lines with large equivalent widths, making the
weak intercombination lines, such as \ion{N}{4}] $\lambda 1486$, \ion{O}{3}]
$\lambda 1664$, and \ion{N}{3}] $\lambda 1750$, easier to detect.  This is
the most comprehensive emission-line abundance analysis so far for a $z > 4$
quasar.

\section{Data \& Analysis}

	The spectrum of BR2248-1242 was obtained at the MMT Observatory with
the Red Spectrograph on September 11, 1994.  The full width at half maximum
(FWHM) spectral resolution is $\sim$10 \AA\, which yields a difference of only
$\sim$10\% between the observed and actual FWHM of even the most narrow
component.  The wavelength range was chosen
to span the redshifted \Lya\ $\lambda 1216$ to \ion{N}{3}] $\lambda 1750$
interval.  This corresponds to an observed wavelength range of $\sim$5500 \AA\
to nearly 10,000 \AA.  The observation and data reduction details are
discussed in Constantin et al. (2001).  Figure 1 shows the observed spectrum.

	We measured the emission lines using tasks in the
IRAF\footnote{The Image Reduction and Analysis Facility (IRAF) is
distributed by the National Optical Astronomy Observatories, which is operated
by the Association of Universities for Research in Astronomy, Inc. (USRA), 
under cooperative agreement with the National Science Foundation.} software
package.  We obtained the Galactic \ion{H}{1} column density (Dickey \& Lockman 1990) and calculated a reddening of $E_{B-V} = 0.0693$.  We then performed a
reddening correction and used the task NFIT1D to fit the continuum with a
powerlaw of the form $F_{\nu}$ $\propto$ $\nu^{\alpha}$.  The fit has a
spectral index, $\alpha =-1.09$ (see Figure 1) and is constrained by the flux
in wavelength intervals between the emission lines, namely 6913--6991 \AA,
8733--8898 \AA, and 9278--9777 \AA.  We next used the task SPECFIT
(Kriss 1994), which employs a $\chi^{2}$ minimization routine, to fit each line
with one or more Gaussian profiles.  Figure 2 shows the fits.  Table 1 lists
the line fluxes and rest-frame equivalent widths (REWs) as measured above
the fitted continuum.  The fluxes, REWs, and FWHMs given in Table 1 are from
the total fitted profiles, which can include several multiplet components and
up to 3 Gaussian profiles per component (see below).  Table 2 lists the rest
wavelength, flux, REW, and FWHM of each individual Gaussian profile.  We
calculate the emission line redshift to be $z=4.156$ based on the narrow
Gaussian profiles fit to \Lya\ and \ion{C}{4}.  We use this redshift to
calculate the rest wavelength for each component. 
  
We fit \ion{C}{4} $\lambda1549$ with three Gaussian profiles, including a
very broad profile (FWHM $\approx 22500$~km~s$^{-1}$).  This broad pedestal
includes the unidentified $\sim$1600 \AA\ emission feature that has been noted
in earlier studies (Laor et al. 1994; Boyle 1990; Wilkes 1984).  We ignore the
doublet splitting in \ion{C}{4} because it is small compared to the observed
line widths.  \ion{N}{4}] $\lambda 1486$, \ion{He}{2} $\lambda 1640$, and
\ion{O}{3}] $\lambda 1664$ are all fit with one Gaussian profile.  Each
component of the \ion{Si}{2} $\lambda\lambda 1527,1533$ and \ion{Si}{4}
$\lambda\lambda 1393,1403$ doublets are fit with one Gaussian profile.  The
fluxes of these components are tied to a 1:1 ratio because this yields a
better fit than a 2:1 ratio.  In every multiplet fit, we fix the ratio of
the central wavelengths to the known ratio of the rest wavelengths and tie
together the FWHMs of each component.  The \ion{N}{3}] $\lambda 1750$ and 
\ion{O}{4}] $\lambda 1403$ multiplets are fit with one Gaussian profile for
each of their 5 components.  The flux component ratios in these cases are
tied to the ratio of the statistical weight $g$ times the $A$ value for each
transition, using atomic data obtained from the National Institute of
Standards and Technology 
(http://aeldata.phy.nist.gov/PhysRefData/contents-atomic.html) and
Nussbaumer \& Storey (1982).
$\Lya$ $\lambda 1216$ is fit with three Gaussian components, with only
the red side used for $\chi^{2}$ minimization because of contamination by
the \Lya\ forest on the blue side.  Each component of the \ion{N}{5}
$\lambda 1240$ doublet is fit with two Gaussian profiles with FWHMs fixed to
the values obtained from the narrow and intermediate width profiles of our
fit to \ion{C}{4}.  This connection between \ion{C}{4} and \ion{N}{5} can
be justified because they are both high ionization lines with similar
excitation and emission properties (Dietrich \& Wilhelm-Erkens 2000).  The 
relative fluxes of the two components of the \ion{N}{5} doublet are set to
2:1 for each component, again because this yields a better fit than a 1:1
ratio.  We fit the \ion{Si}{2} $\lambda \lambda 1260,1264,1265$ triplet with
one Gaussian profile for each of the three components.  Because the flux of
\ion{Si}{2} is so much smaller than $\Lya$ and \ion{N}{5}, \ion{Si}{2}
is not well defined by the $\chi^{2}$ minimization.

 	The primary uncertainty in our flux measurements is the continuum 
location.  Many of the lines are blended, which introduces more uncertainty,
particularly for \ion{N}{5} in the wing of $\Lya.$  The broad component in
the \ion{C}{4} fit introduces another uncertainty because it includes the
$\sim$1600 \AA\ feature.  We do not include this broad component in our
calculated of flux ratios (\S3\ below) because it is believed to be an
unrelated emission feature (Laor et al. 1994).  We estimate the one $\sigma$
standard deviation of our measurement of the flux of $\Lya$ to be $\sim$10\%
based on repeated measurements with the continuum drawn at different levels.
By the same method, we estimate the uncertainty in \ion{C}{4}, \ion{N}{5},
\ion{He}{2}, and \ion{O}{3}] to be $\sim$15--20\% and the uncertainty in the
remaining lines with  REW $\leq 10$ \AA\ to be $\sim$25\%. 

\section{Results}

        Table 3 lists various measured flux ratios and the metallicities
inferred from comparisons to the theoretical results from Hamann et
al. (2001).  Our preferred estimates of the metallicity are obtained from
the model in Hamann et al. that uses a segmented powerlaw for the
photoionizing continuum shape (see Figure 3).  This continuum shape is a
good approximation to the average observed continuum in quasars (Zheng et
al. 1997; Laor et al.  1997).  The metallicity ranges in Table 3 represent
the range of results obtained by comparisons to all three different continuum
shapes calculated by Hamann et al. and shown in Figure 3.  These theoretical
uncertainties are in addition to any uncertainty in the measured line
strengths (\S2).

	All of the line ratios in Table 3 yield a metallicity of $Z\approx$
1--3 \Zsun.  From a theoretical viewpoint, \ion{N}{3}]/\ion{O}{3}] is the most
reliable of the intercombination line ratios (Hamann et al. 2001).
\ion{N}{5}/\ion{He}{2} is also a useful ratio, whereas \ion{N}{4}]/\ion{O}{3}]
and \ion{N}{4}]/\ion{C}{4} can be unreliable because they are more sensitive
to non-abundance effects and generally yield a wider range of results than
\ion{N}{3}]/\ion{O}{3}].  Based on which ratios are most accurately measured,
and which are most reliable from a theoretical viewpoint (Hamann et al. 2001),
we estimate the overall metallicity of BR2248-1242 to be roughly $Z\approx$
2 \Zsun.

	It is important to note, however, that the actual metallicity of
BR2248-1242 may be 2--3 times higher than $Z\approx$ 2 \Zsun\ because the
theoretical models we use assume $q\approx0$ in Equation 1.  For environments
with rapid chemical evolution, such as massive galactic nuclei,
$q\approx0.2$--0.5 may be more appropriate (Hamann et al. 2001).  Therefore,
the metallicity of BR2248-1242 is broadly consistent with previous emission
line results of $Z\approx$ 1--9 \Zsun \ derived for other quasars (Hamann et
al. 2001; Dietrich \& Wilhelm-Erkens 2000; Hamann \& Ferland 1999; Korista et
al. 1998; Ferland et al. 1996).  

For any appropriate $q$ value, BR2248-1242 may still yield a lower bound on
the metallicity of luminous quasars because it appears to have an unusually low
black hole mass (and so may reside in a lower mass, lower metallicity galaxy).
We estimate the black hole mass in BR2248-1242 to be $M_{BH} \approx 4.77
\times 10^{8} \ \Msun$, based on the measured FWHM of \ion{C}{4} and the
formulae given in Kaspi et al. (2000)
\begin{equation}
R_{\rm BLR} = (32.9^{+2.0}_{-1.9}) \left[ {{\lambda L_{\lambda}(5100 {\rm \AA})} \over {10^{44} \ {\rm ergs \ s^{-1}}}} \right] ^{0.700 \pm 0.033} \ {\rm lt-days}
\end{equation}
\begin{equation}
M = 1.464 \times 10^{5} \left( {{R_{{\rm BLR}}} \over {{\rm lt-days}}} \right) \left( {{v_{{\rm FWHM}}} \over {{\rm 10^{3}\ km\ s^{-1}}}} \right) ^{2} \ \Msun
\end{equation}
where $R_{\rm BLR}$ is the radius of the broad line region for $H \beta$, and
$R_{\rm BLR}$(\ion{C}{4}) $\approx$ 0.5 $R_{\rm BLR}(H \beta)$ is the radius
of the \ion{C}{4} region (Peterson 2001).  The FWHM of \ion{C}{4} (1620 km
${\rm s}^{-1}$) is represented by $v_{\rm FWHM}$ and $M$ is the mass of the
black hole.  We derive $\lambda L_{\lambda}(1450 {\rm \AA})$ from the
observed flux (Dietrich et al. 2002 in preparation) and use a powerlaw of
the form $F_{\nu}$ $\propto$ $\nu^{\alpha}$ with $\alpha = -0.4$ to
approximate the continuum shape and estimate $\lambda L_{\lambda}(5100
{\rm \AA})$.  We select $\alpha = -0.4$ based on average quasar spectra
from Vanden Berk et al. (2001) and Dietrich et al. (2002 in preparation).
The derived black hole mass in BR2248-1242 is unusually low because of its
unusually narrow emission lines.  Previous studies suggest a correlation
between black hole mass and the overall bulge/spheroidal component mass of
the surrounding galaxy (Gebhardt et al. 2001; Ferrarese \& Meritt 2000;
Laor 2001; Wandel 1999).  Together with the  well known relationship between 
the mass and metallicity of galaxies (Faber 1973; Zaritsky et al. 1994;
Jablonka et al. 1996), these results predict a relationship between the mass
of the black hole and the metallicity surrounding the quasar
(Warner et al. 2002 in preparation).  BR2248-1242 appears to have a low black
hole mass, and so may be expected to have relatively low metal abundances.

\section{Discussion}

	The result for $Z\gtrsim \Zsun$ in BR2248-1242 suggests that an
episode of rapid and extensive star formation occurred before the observed
$z=4.16$ epoch.  The spectral similarity of BR2248-1242 to other $z > 4$
quasars (apart from the narrow line widths, see Constantin et al. 2001)
suggests further that the resulti for $Z\gtrsim \Zsun$ should apply generally.
The time available for the chemical enrichment at $z=4.16$ is only $\sim$1.5
Gyr (for H$_{0}=65$ km s$^{-1}$ Mpc$^{-1}$, $\Omega_{m}=0.3$ and
$\Omega_{\Lambda} \lesssim 1$).  This star formation episode early in the
history of the universe would be consistent with observations showing that
the central regions of massive galaxies today contain old, metal rich stars
(McWilliam \& Rich 1994; Bruzual et al. 1997; Worthey et al. 1992; Idiart et
al. 1996).  This high redshift evolution would also be well within the
parameters derived in some recent simulations, which show that the densest
proto-galactic condensations can form stars and reach solar or higher
metallicities at $z\gtrsim6$ (Ostriker \& Gnedin 1997; Haiman \& Loeb, 2001).
Recent observations of the highest redshift quasars suggest that reionization
occurred around $z\sim6$ (Becker et al. 2001; Djorgovski et al.  2001).  If
the reionization is due to stars, this would provide further evidence for
substantial star formation beginning at $z\gtrsim6$.  Quasar abundance
studies can provide observational tests of these evolution scenarios.  The
next step is clearly to extend the abundance analysis to the
highest possible redshifts, and to compare the results across wide ranges in
redshift, host galaxy types, and central black hole mass.

\noindent {\it Acknowledgements:} We are grateful to Matthias Dietrich,
Bassem Sabra, Eric McKenzie, and Catherine Garland for their comments
on this manuscript.  We acknowledge financial support from the NSF
via grants AST99-84040 and AST98-03072.

\newpage
\begin{figure}
\vbox{
\centerline{
\psfig{figure=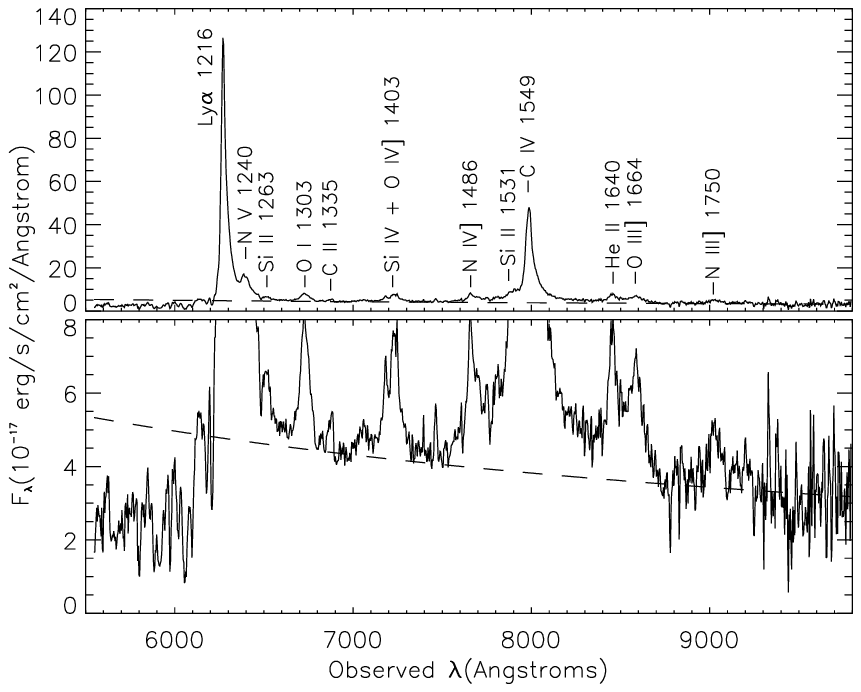}
}
\vskip 14pt
\vskip 14pt
\caption{
Spectrum of BR2248-1242 with our fit to the continuum overplotted
(dashed line).  The spectrum has been corrected for extinction.}
}
\end{figure}

\begin{figure}
\vbox{
\centerline{
\psfig{figure=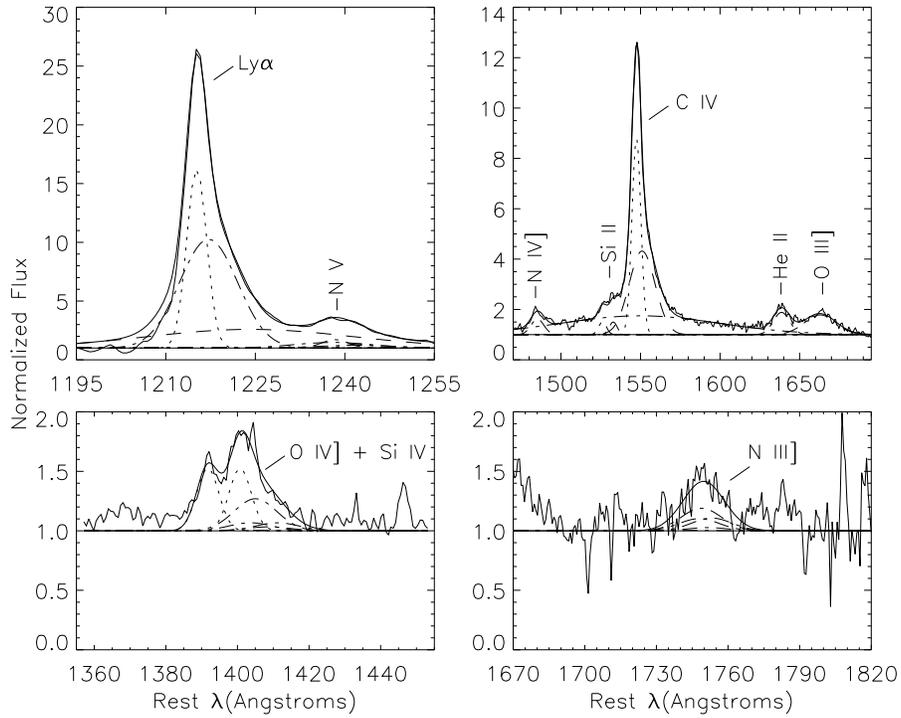}
}
\vskip 10pt
\caption{
Gaussian fits of emission lines of BR2248-1242.  The continuum is normalized
by the power law fit shown in Figure 1.  The continuum (at unity) and
composite line fits are plotted as solid curves.  The individual components
of the fits are shown as dotted and dashed lines.  Larger versions of these
plots are available at http://www.astro.ufl.edu/$\sim$warner/BR2248/.}
}
\end{figure}

\begin{figure}
\vbox{
\centerline{
\psfig{figure=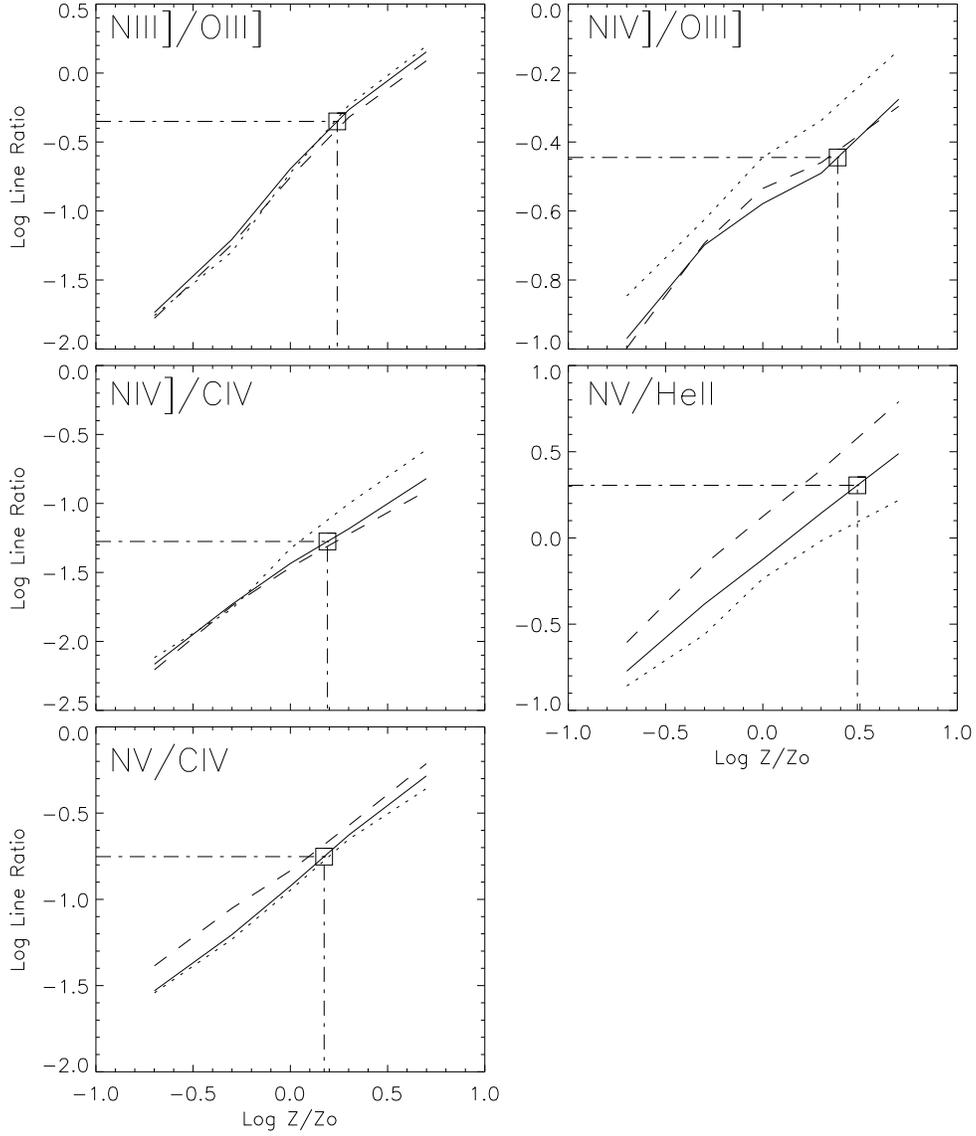}
}
\vskip 14pt
\caption
{
Comparison of flux ratios measured in BR2248-1242 to theoretical results from
Figure 5 of Hamann et al. (2001).  Comparisons of our measured line
ratios to the theoretical model represented by the solid line yield our
preferred estimates for metallicity. 
The solid line is obtained from the model in Hamann et al. that uses a
segmented powerlaw for the photoionizing continuum shape, which is a good
approximation to the average observed continuum in quasars.  The dotted
line corresponds to the model that uses an incident spectrum
defined by Matthews \& Ferland (1987), while the dashed line corresponds to
a powerlaw with index $\alpha = -1.0$ across the infrared through X-rays .}
}
\end{figure}

\begin{table}
\begin{center}
\caption{Emission Line Data}
\vspace*{0.1in}
\begin{tabular}{lcccccc}
\tableline
\tableline
\noalign{\vskip 4pt}
Line & $\lambda_{obs}^{a}$ & Flux$^{b}$ & Flux/$\Lya$ & REW${^a}$ & FWHM$^{c}$\\
\noalign{\vskip 4pt}
\tableline
$\Lya$ $\lambda 1216$ & 6270.3 & 6040 & 1.000 & 246 & 1450\\
\ion{N}{5} $\lambda 1240$ & 6396.6 & 420 & 0.070 & 18 & 2380\\
\ion{Si}{2} $\lambda 1263$ & 6509.9 & 90 & 0.014 & 4 & 3660\\
\ion{O}{1} $\lambda 1303$ & 6728.7 & 250 & 0.041 & 10 & 3020\\
\ion{C}{2} $\lambda 1335$ & 6874.2 & 30 & 0.005 & 1 & 1270\\
\ion{Si}{4} $\lambda 1397$ & 7204.5 & 180 & 0.029 & 8 & 3490\\
\ion{O}{4}] $\lambda 1403$ & 7254.8 & 140 & 0.023 & 7 & 3240\\
\ion{N}{4}] $\lambda 1486$ & 7663.9 & 130 & 0.021 & 6 & 1910\\
\ion{Si}{2} $\lambda 1531$ & 7892.1 & 120 & 0.020 & 6 & 2450\\
\ion{C}{4}  $\lambda 1549$ & 7984.8 & 2400 & 0.398 & 122 & 1620\\
\ion{He}{2} $\lambda 1640$ & 8455.3 & 210 & 0.035 & 11 & 2200\\
\ion{O}{3}] $\lambda 1665$ & 8580.6 & 350 & 0.059 & 19 & 4200\\
\ion{N}{3}] $\lambda 1751$ & 9029.2 & 160 & 0.026 & 9 & 3530\\

\tableline
\end{tabular}
\end{center}
$^a$ In units of \AA\\
$^b$ In units of 10$^{-17}$ ergs~cm$^{-2}$~s$^{-1}$\\ 
$^c$ In units of km~s$^{-1}$\\ 
\end{table}

\begin{table}
\begin{center}
\caption{Emission Line Components}
\vspace*{0.1in}
\begin{tabular}{lcccc}
\tableline
\tableline
\noalign{\vskip 4pt}
Component & $\lambda_{rest}^{a}$ & Flux$^{b}$ & REW$^{a}$ & FWHM$^{c}$\\
\noalign{\vskip 4pt}
\tableline
$\Lya$ 1   & 1216.1 & 1580 & 64 & 970\\
$\Lya$ 2   & 1218.1 & 2780 & 112 & 2810\\
$\Lya$ 3   & 1225.1 & 1690 & 69 & 10020\\
\ion{N}{5} 1  & 1240.2 & 140 & 6 & 1280\\
\ion{N}{5} 2  & 1240.6 & 280 & 12 & 3710\\
\ion{Si}{2}   & 1262.8 & 90 & 4 & 3390\\
\ion{O}{1}   & 1305.0 & 250 & 10 & 3020\\
\ion{C}{2}   & 1333.2 & 30 & 1 & 1230\\
\ion{Si}{4}  & 1397.3 & 180 & 8 & 1570\\
\ion{O}{4}]  & 1407.1 & 140 & 7 & 3000\\
\ion{N}{4}]  & 1486.3 & 130 & 6 & 1910\\
\ion{Si}{2}  & 1530.7 & 120 & 6 & 1140\\
\ion{C}{4} 1 & 1548.6 & 1070 & 55 & 1280\\
\ion{C}{4} 2 & 1552.2 & 1330 & 68 & 3710\\ 
\ion{C}{4} 3 & 1551.0 & 1820 & 93 & 22570\\
\ion{He}{2}  & 1639.9 & 210 & 11 & 2200\\
\ion{O}{3}]  & 1664.2 & 350 & 19 & 4200\\
\ion{N}{3}]  & 1751.2 & 160 & 9 & 3390\\
\tableline
\end{tabular}
\end{center}
$^a$ In units of \AA\\
$^b$ In units of 10$^{-17}$ ergs~cm$^{-2}$~s$^{-1}$\\
$^c$ In units of km~s$^{-1}$\\
\end{table}

\begin{table}
\begin{center}
\caption{Emission Line Ratios}
\vspace*{0.1in}
\begin{tabular}{lcc}
\tableline
\tableline
\noalign{\vskip 4pt}
Lines & Ratio & Z/$\Zsun$\\
\noalign{\vskip 4pt}
\tableline
\noalign{\vskip 4pt}
\ion{N}{3}]/\ion{O}{3}] & 0.45 & 1.74$^{+0.16}_{-0.05}$\\
\noalign{\vskip 4pt}
\ion{N}{4}]/\ion{O}{3}] & 0.36 & 2.43$^{+0.00}_{-1.20}$\\
\noalign{\vskip 4pt}
\ion{N}{4}]/\ion{C}{4} & 0.05 & 1.55$^{+0.16}_{-0.44}$\\
\noalign{\vskip 4pt}
\ion{N}{5}/\ion{He}{2} & 2.02 & 3.07$^{+3.94}_{-1.48}$\\
\noalign{\vskip 4pt}
\ion{N}{5}/\ion{C}{4} & 0.18 & 1.49$^{+0.09}_{-0.25}$\\
\noalign{\vskip 4pt}
\tableline
\end{tabular}
\end{center}
\end{table}

\end{document}